\documentclass{article}
\usepackage{arxiv}
\usepackage[T1]{fontenc}
\usepackage{comment}
\usepackage{booktabs}
\usepackage{multirow}
\usepackage{amsfonts}
\usepackage{caption}
\usepackage{subcaption}

\usepackage{graphicx}
\usepackage{enumitem}
\usepackage{titlesec}

\usepackage{wrapfig}
\usepackage{rotating}
\usepackage{amsmath}

\DeclareMathOperator*{\argmax}{arg\,max}

\usepackage{color}
\PassOptionsToPackage{hyphens}{url}\usepackage{hyperref}

\begin{document}

\renewcommand{\headeright}{Accepted at SECAI Workshop / ESORICS 2023}
\renewcommand{\undertitle}{(accepted SECAI/ESORICS 2023~--~best paper)}
\renewcommand{\shorttitle}{}

\title{Fault Injection and Safe-Error Attack for Extraction of Embedded Neural Network Models}

\date{}


\author{ {Kevin Hector, Pierre-Alain Moëllic, Mathieu Dumont} \\
	CEA Tech, Centre CMP, Equipe Commune CEA Tech - Mines Saint-Etienne, F-13541 Gardanne, France \\
	Univ. Grenoble Alpes, CEA, Leti, F-38000 Grenoble, France\\
	\texttt{\{name\}.\{surname\}@cea.fr} \\
	\And
	{Jean-Max Dutertre} \\
	Mines Saint-Etienne, CEA, Leti, Centre CMP,\\ F-13541 Gardanne France\\
	\texttt{dutertre@emse.fr} \\
}

\maketitle              

\begin{abstract}
Model extraction emerges as a critical security threat with attack vectors exploiting both algorithmic and implementation-based approaches. 
The main goal of an attacker is to steal as much information as possible about a protected victim model, so that he can mimic it with a substitute model, even with a limited access to similar training data. Recently, physical attacks such as fault injection have shown worrying efficiency against the integrity and confidentiality of embedded models. We focus on embedded deep neural network models on 32-bit microcontrollers, a widespread family of hardware platforms in IoT, and the use of a standard fault injection strategy~--~Safe Error Attack (SEA)~--~to perform a model extraction attack with an adversary having a limited access to training data. Since the attack strongly depends on the input queries, we propose a black-box approach to craft a successful attack set.
For a classical convolutional neural network, we successfully recover at least 90\% of the most significant bits with about 1500 crafted inputs. These information enable to efficiently train a substitute model, with only 8\% of the training dataset, that reaches high fidelity and near identical accuracy level than the victim model.    

\end{abstract}


\section{Introduction}
\label{introduction}
Deep neural network models suffer from many critical security issues including confidentiality and privacy threats. A growing concern is \textit{model extraction} attacks that, basically, aim at stealing a victim model. Different adversarial goals have to be distinguished~\cite{jagielski_high_2020}. First, an adversary may want to clone the model (\textit{fidelity} objective) and build a \textit{substitute} model with an architecture and parameters as close as possible to the victim model. Thus, several works (as ours) concerned the challenging task of the extraction of model parameters. A second objective (\textit{task accuracy} objective) is to efficiently steal the performance of the victim model (to reach equal or better performance at lower cost) \cite{orekondy2019knockoff,tramer_stealing_2016}. For that case, having similar architecture or parameter values is not compulsory.

This work focuses on a model parameters extraction with \textit{fidelity} objective. This challenge receives a growing interest by the exploitation of very different attack vectors, from cryptanalysis-based methods as in~\cite{carlini_cryptanalytic_2020,rolnick_reverse_engineering_2020} to learning-based techniques focused on the substitute model~\cite{papernot_practical_2017,tramer_stealing_2016}. Because of very limited knowledge on the training distribution, both approaches need a very large amount of queries to the victim model that could be prohibitive in many attack context. 
However, the large-scale deployment of models in a wide variety of hardware platforms fosters the emergence of new attack vectors such as \textit{side-channel} (SCA)~\cite{joud2023practical} and \textit{fault injection} analysis (FIA)~\cite{deepsteal2022}.

For now, the use of fault injection techniques for model extraction has only been performed with rowhammer~\cite{kim2014flipping}, such as in~\cite{deepsteal2022} that leverage hardware flaws of DRAM to partially guess parameter values. However, such rowhammer-based approaches    exclude other platforms but CPU-based ones with DRAM memory and are known to be complex to carry out in practice. Thus, our objective is to widen the scope of FIA-based approaches for model extraction by focusing on models embedded in constrained devices with the widespread 32-bit microcontroller platforms, massively used for IoT applications. For such devices, bit-flips with rowhammer as in~\cite{deepsteal2022}, are not practicable since they rely on Flash memory. Our work is the first to demonstrate that a well-known attack strategy against cryptographic modules is possible and can reach consistent results regarding the state-of-the-art. Our contributions are the following:
\begin{itemize}[nosep]
    \item We demonstrate a new extraction method based on a Safe-Error Attack (SEA) that exploits a bit-set fault model ($\texttt{0}\rightarrow\texttt{1}$, $\texttt{1}\rightarrow\texttt{1}$). SEA is a standard FIA strategy that relies on a simple but powerful principle: injecting a fault on a secret parameter of a program may or may not lead to a faulted output according to the parameter value.       
    \item SEA enables to recover the most significant bit values of the victim model parameters. These information enable to efficiently constrain the substitute model training, even with limited training data.
    \item We use full 8-bit quantized models
    as for real-world embedded applications, a setting that makes our approach more challenging.
    \item We show that the model inputs significantly impact extraction performance and we propose an input generation method to increase the recovery rate. 
\end{itemize}

\noindent\textbf{Availability.} Data and codes related to our work are publicly available on \url{https://gitlab.emse.fr/securityml/lfi_sea_modelextraction}.

\section{Background}
\label{backgrounds}
\subsection{Model extraction}
\label{model_extraction}
Threats against the confidentiality of ML models, especially deep neural networks, have been extensively studied with both algorithmic and implementation-based attacks including physical attacks (especially side-channel analysis). 

Let's consider a supervised neural network model $M_W$, with $W$ its internal parameters. The input domain is $\mathcal{X}$ and $M$ is trained thanks to a training dataset $(X^{train}, Y^{train})$, with $Y^{train} \subset \mathbb{R}^{K}$ the set of labels ($K$ is the number of labels). For an input $x \in \mathcal{X}$, the output prediction is $M_W(x) \in \mathbb{R}^{K}$ and the predicted label is $\hat{y} = \argmax(M_W(x))$. In a \textit{fidelity} scenario, the goal of an adversary is to craft a model $M'_{\Theta}$ that mimic $M_W$ as perfectly as possible regarding his knowledge and ability. Note that $M'$ aims at providing the same predictions as $M$, including potential mistakes from $M_W$.  Obviously, this goal strongly depends on how the \textit{similarity} between two models is defined. A typical approach~\cite{jagielski_high_2020} is to measure the agreement of both models at the label-level, i.e. a classical objective is to have $\argmax(M'_\Theta(x)) = \argmax(M_W(x))$ for every $x$ sampled from a target distribution over $\mathcal{X}$. A more complex and optimal extraction, \textit{Functionally Equivalent Extraction}, aims to reach $M'(x)=M(x), \forall x\in \mathcal{X}$. Note that the strongest possible attack leading to a substitute model with exactly the same architecture and same parameters ($W=\Theta$) is infeasible by only exploiting input/output pairs from the victim model~\cite{jagielski_high_2020}.

A first type of methods relies on the training of $M'_\Theta$ with several works based on \textit{active learning} principles. The challenge relies on the classical assumption that an adversary has a very limited (if any) access to the original training data. Therefore, a critical limitation is the need of a very large amount of query/output pairs, collected from the victim model, to build an efficient substitute training dataset.
A second approach is based on a full mathematical recovery by exploiting chosen input/output pairs as well as gradient-based properties of the model, such as the critical point ($x=0$) of the second derivatives of ReLU in~\cite{carlini_cryptanalytic_2020}. The state-of-the-art result of this approach is a near perfect recovery (worst case error of $2^{-25}$) of the 100K parameters of a 2-layer MLP using $2^{21.5}$ queries. The attack has not been scaled to deeper models.       

The last methods leverage the power of physical attacks to exploit information leakages, regularly demonstrated in many security applications such as cryptography~\cite{Barenghi2012}. More essentially, side-channel analysis have been proposed to guess the values of a model parameters with timing analysis~\cite{maji2021leaky} or typical correlation power analysis~\cite{batina2019csi} even if the task still gathers critical open challenges~\cite{joud2023practical} for a full extraction of real-world embedded models. 

\subsection{Fault Injection Attacks}
\label{dfa}
Fault injection attacks (FIA) are active hardware threats that consist in faulting the operations of a target circuit for the purpose of extracting a secret or gaining an unauthorized access~\cite{Barenghi2012}. They generally require a physical access to the target~\cite{breier2022practical} but some variants can be carried out remotely like rowhammer \cite{kim2014flipping} that targets DRAM. By heavily addressing
some memory rows, an attacker disturbs the adjacent rows (in the victim's address space) which may typically leads to bit-flip faults. Other remote approaches used software breaches in dynamic voltage and frequency scaling modules~\cite{qiu2019voltjockey}.
\\

\noindent\textbf{Laser Fault Injection.} 
Typical injection means gather low-cost techniques such as voltage or clock glitching that globally (i.e., spatially) alter the target device and moderate/high-cost methods such as electromagnetic pulse or laser beam (usually, near-infrared). Laser fault injection (LFI) is particularly used in security testing centers and for certification purpose because it enables powerful analysis with high temporal and spatial accuracy~\cite{agoyan2010flip}. 
LFI makes it possible to inject faults at bit-level according a data-dependent fault model\footnote{Note that we use the term \textit{fault model} in a restrictive way to describe the mathematical properties of the fault injection process.}. Usually, a bit-flip data-independent fault model is used to model FIAs, according to which a faulted bit is inverted whatever its original value (the bit is said to be \textit{flipped}: $\texttt{0}\rightarrow\texttt{1}$, $\texttt{1}\rightarrow\texttt{0}$).
Using LFI, the fault model be either a bit-set or a bit-reset. When a bit-set fault is injected, an actual error is induced when the initial bit value was $\texttt{0}$, the faulted bit then switches to $\texttt{1}$  ($\texttt{0}\rightarrow\texttt{1}$). When the original bit value is already at $\texttt{1}$, it stays at $\texttt{1}$ ($\texttt{1}\rightarrow\texttt{1}$): the targeted bit is \textit{safe} from any error. This fault model is said to be \textit{data-dependent} as the injection of a fault depends on its initial data state. A bit-reset fault model is linked to a symmetric behavior ($\texttt{0}\rightarrow\texttt{0}$, $\texttt{1}\rightarrow\texttt{0}$).
The physical phenomenon of a bit-set fault model with a laser shot in a floating gate transistor relies on the creation of a photolectric current that induces voltage transients allowing to perform a fault injection~\cite{colombier2019laser}.
\\

\noindent\textbf{Safe-Error Attack}. Data-dependent fault model provides additional information to an attacker by simply observing the error-free response of the target. That is the basic principle of the Safe-Error Attack (SEA) described in~\cite{yen2000checking}. SEA relies on the observation that a fault could lead or not to an incorrect output depending on a secret data. In our case, the secret is a parameter of the victim model and the output is its predictions. Since SEA has been mostly applied for cryptography, we illustrate the attack principle with a secret key recovering task for an encryption algorithm that outputs a ciphertext from an input plaintext. A bit-set is injected on the first bit of the secret key. If the obtained ciphertext is erroneous, the key bit is $\texttt{0}$ (an actual $\texttt{0}\rightarrow\texttt{1}$ fault was injected), if it is error-free, the key bit is $\texttt{1}$ (because the key bit was actually unfaulted $\texttt{1}\rightarrow\texttt{1}$). Then, the whole key can leak with an iterative attack on all its bits. With this work, we show that the SEA is also relevant to extract information from a DNN as described in section \ref{sea}.
\\

\noindent\textbf{Fault injection on 32-bit microcontrollers.} LFI in SRAM or D flip-flop memories follows a data-dependent fault model. Well-defined and precise locations of these memory cells yield either a bit-set or a bit-reset fault-model as assessed on experimental basis \cite{DUT18fdtc,ROS13sram}. The challenge is to find with certainty the points of interest of many memory bits involved in an SEA which may question the practical feasibility of SEA while targeting SRAM cells and D flip-flops.
However, microcontrollers store their program and data (such as a DNN model) in embedded Flash memories (the most usual kind of embedded non-volatile memory).
At read time, these data are sensitive to LFI according a bit-set fault model when read from the Flash memory for 32-bit microcontroller targets as reported by \cite{colombier2019laser,menu2020single}. The experiments carried out in these reference works showed that it is feasible: (1) to achieve a 100\% success rate when inducing a bit-set, and (2) to chose at will the index of a single bit to be faulted among the 32 bits of the read data, thanks to the regular and orderly architecture of an embedded Flash (which generally follows a NOR architecture). Hence, the experimental state-of-the-art shows that performing an SEA using LFI on the Flash memory of a microcontroller is within the reach of attackers who can access a LFI setup.

\section{Related Works}
\label{related_works}

Few works addressed the extraction of ML models by using fault injection. In \cite{breier2021sniff}, Breier \textit{et al.} proposed to extract the parameters of the last layer only of a victim model. This work sets in a restricted scenario where the adversary perfectly knows everything of the victim model except the parameters of the very last output layer. The authors claimed that it is the case in a transfer learning scenario with all the other layers coming from a public pretrained model. To reverse the last layer, they used fault injections to alter the sign of the parameters and demonstrated, by simulations, that only $mn$ faults and $2mn$ executions of the victim within the weighted sum of each neuron are necessary to extract the full weight matrix of the target layer (with $m$ and $n$ the number of the neurons of the last and penultimate layers respectively).  
Our work is significantly different since we target all the parameters of the model and do not use fault injections to mathematically reverse a layer computations (in~\cite{breier2021sniff} a Softmax-layer) but to extract as much information as possible to efficiently train a substitute model.   

Therefore, our main reference is DeepSteal \cite{deepsteal2022} 
 with which we share the objective and threat model. DeepSteal exclusively concerns DRAM platform (in~\cite{deepsteal2022}, Intel i5 CPU) and leverages bit-flip faults with a rowhammer attack~\cite{kwong2020rambleed} to recover MSB of a victim model parameters.
 Then, the authors propose to use the MSB recovered to constraint the training of a substitute model. For the parameters with recovered bits, a range of possible values and a mean value are defined. At training time, these mean values act as a classical weight-penalty regularization (see Section~\ref{target_training} for details). Our work aims at demonstrating that this two-step methodology is actually generalizable to another type of platforms, i.e. 32-bit microcontrollers, with a different fault model (bit-set) and exploitation methods (SEA and input crafting). Our approach is suitable to any fault injection means that lead to data-dependent faults (bit-set or bit-reset).

\section{Threat Model}
\label{threat_model}
Our work is positioned in a context where an attacker tries to steal parameters of a model in order to copy it or to prepare future attacks.
Our adversary targets embedded neural network models in platform such as 32-bit microcontrollers, thus 8-bit quantized models specifically designed for embedded inference.

We set in a traditional grey-box context for model extraction~\cite{carlini_cryptanalytic_2020,deepsteal2022}, with an adversary knowing the victim model architecture but not its parameter values and accesses to less than 10\% of the training dataset. In some cases, model's architecture is effectively already known, easy to guess or previously extract with an appropriate attack (including physical ones such as~\cite{batina2019csi}). Likewise, limited access to training data also corresponds to real-world applications without publicly available benchmarks, as studied in many active learning-based extraction methods~\cite{barbalau2020black,chandrasekaran2020exploring,orekondy2019knockoff,papernot_practical_2017}.    

The adversarial ability is basically twofold. First, the adversary has an unlimited black-box access to the model by querying it and getting the (normalized) outputs (i.e., not the \textit{logits}). Importantly, working with full quantized models (8-bit), the available prediction scores are also quantized. Second, the adversary has a fault injection means that can yield a data-dependent fault model (e.g. a laser setup) and a clone device on which the attacker can profile the Flash memory layout to accurately control the fault injection process. This profiling process does not need the target inference program but only simple read/write memory procedures, as in~\cite{colombier2019laser}.
\\

\noindent\textbf{Notations.} The victim neural network is noted $M_W$ with $W$ its parameters. The substitute model is $M^{s}$ with parameters $W_{s}$. $M$ performs a classification task with inputs $x \in \mathcal{X} (\subset \mathbb{R}^{d}$) and the output predictions $M_{W}(x) \in \mathbb{R}^{K}$ with $\mathcal{X}$ the input domain and $K$ the number of labels. Then, the predicted label is $y=\argmax(M_W(x))$ and the correct label is noted as $y^{*}$. A set of inputs and predictions are respectively noted as $X$ and $Y$. The loss function is the categorical cross-entropy simply noted $\mathcal{L}_{CE}$. Our method is based on a safe-error attack with bit-set faults perform on the victim model parameters. The faulted parameters are referred as $\widetilde{W}$, then $M_{\widetilde{W}}$ is the resulting faulted model and $\widetilde{Y}$ is a set of faulted predictions. Our models are 8-bit quantified with signed integers (two's complement representation): a parameter $w$ is represented as $b_{0}b_{1}...b_{7}$ with $b_{i}$ a bit value and $b_0$ the Most Significant Bit (hereafter, MSB).

\section{Experimental setup}
\label{setup}
\subsection{8-bit quantized neural network models}
Our work is focused on 8-bit inference implementations that correspond to real-world applications using constrained embedded platforms. 8-bit quantization is the \textit{de facto} practice for embedded models on microcontrollers and is the default configuration in many deployment tools (e.g., NNoM, TF-Lite, CubeMX.AI, MCUNet).
However, the extraction method being based on a SEA, the outputs quantization has a strong impact on the extraction process. 

A first approach is a lite post-training 8-bit quantization of the parameters only for memory footprint purpose: at inference time, the 8-bit stored parameters are then scaled to full-precision values, the computations, activation outputs as well as the prediction scores are in full-precision. 
Because of its easiness, this quantization scheme is used in many simulation works. However, we claim that it may represent a strong limitation and drawbacks since it does not represent the real behavior of embedded models with real-world deployment platforms relying on more complex quantization schemes.  
If we consider this naive lite-quantization, our SEA-based approach straightforwardly extract most of the bits of the victim model parameters. However, the extraction is more challenging when dealing with a full-quantization process that includes the parameters, the activation values and the output prediction scores.

We developed a Python framework based on Pytorch to perform all our simulations with this 8-bit quantization schema. We chose NNoM (Neural Network on Microcontroller)\footnote{https://majianjia.github.io/nnom/} as model deployment library that uses the reference CMSIS-NN library from ARM~\cite{cmsisNN_CNN} as backend. NNoM is open-source with a full access to the C code and allows 8-bit quantization for weights, biases and activation function with a uniform symmetric powers-of-two quantization scheme (as in CMSIS-NN). This scheme is popular for embedded platforms because intrinsic calculations require no division only integer additions, multiplications and bit shifting (see Appendix for details). Our Python framework provides the same outputs (at layer and model-level) than ones provided by NNoM on a Cortex-M platforms. More particularly, NNoM deals with signed 8-bit integers only and does not scale the prediction scores ($\in \mathbb{R}^{+}$) in $[0, 255]$ but in $[0, 127]$.

The powers-of-two quantization scheme is as follow: $x_{i} = \left \lfloor{x_{f}\cdot 2^{7-dec}}\right \rceil$, $dec = \left \lceil{log_{2}\big(max\big(|X_{f}|\big)\big)}\right \rceil$. $X_{f}$ is a 32-bit floating point tensor, $x_{f}$ a value of $X_f$, $x_{i}$ its 8-bit counterpart and $2^{dec}$ the quantization scale.

\subsection{Models and datasets}
We used two classical model architectures. Our first model is a multilayer perceptron composed of three fully-connected layers (128 - 64 - 10 neurons, no bias) with ReLU as activation function. In the rest of the paper, this model is simply referred as \textbf{MLP}. MLP is trained on MNIST, composed of 70k grayscale images ($28\times28$) of digits. The second model is a convolutional neural network composed of three convolutional layers and one fully-connected layer (no bias). This model is a usual reference for embedded models in microcontrollers presented in~\cite{cmsisNN_CNN}. For the rest of this paper, this model is referred as \textbf{CNN}. CNN is trained on the Cifar-10 dataset composed of 60k color images ($32\times32$) among 10 categories. Table~\ref{model_architecture} details our models.

\begin{table}[t!]
\centering
\caption{MLP (MNIST) and CNN (Cifar-10) architecture. We follow the PyTorch naming for fully-connected layers with "Linear". Convolutional layers are composed by 5x5 kernels and followed by a pooling layer (average 2x2).}
\label{model_architecture}
\begin{tabular}{cccc}
\toprule
Layer & \# param & Layer & \# param  \\
\midrule
Inputs (784)& & Inputs (32,32) & \\
Linear 1 (128 neurons), ReLU & 100352 & Conv1 (32 kernels), ReLU & 2400  \\
Linear 2 (64 neurons), ReLU & 8192 & Conv2 (32 kernels), ReLU & 25600  \\
Linear 3 (10 neurons), Softmax & 640 & Conv3 (64 kernels), ReLU & 51200  \\
 & &   Linear (10 neurons), Softmax & 10240  \\
\midrule
MLP & 109184 & CNN & 89440\\
\bottomrule
\end{tabular}
\end{table}

\section{Model Extraction with SEA}
\label{extraction}

\subsection{Overview}
\label{overview}
\begin{figure}[h!]
\centering
  \includegraphics[width=\textwidth]{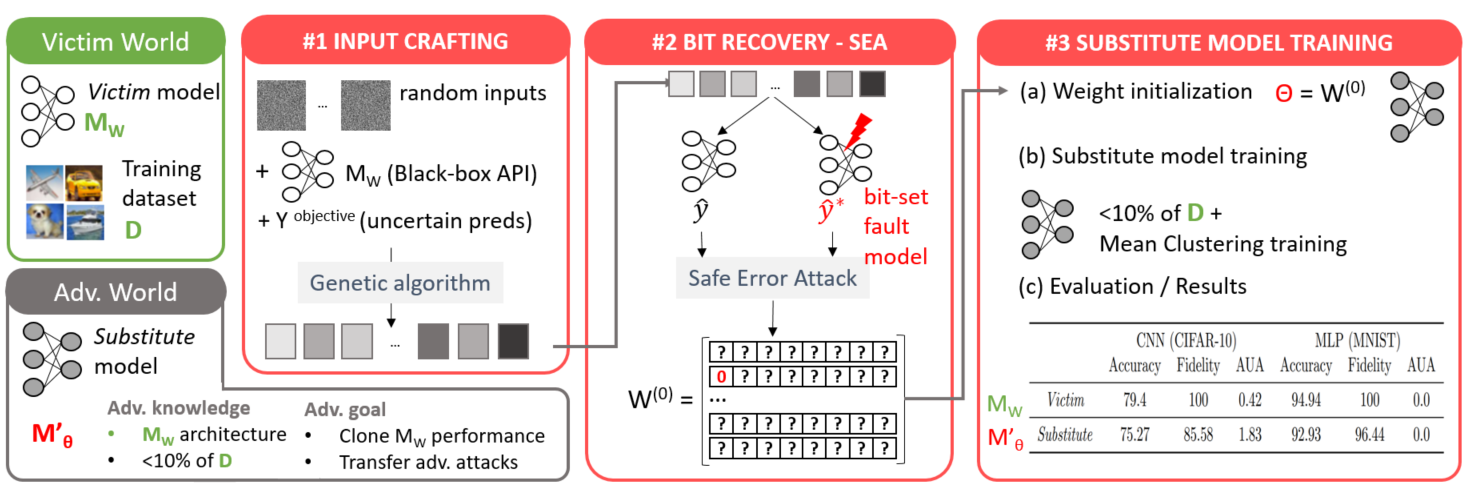}
  \caption{The adversary crafts inputs and performs a safe-error attack exploiting faulted predictions with bit-set fault injections on the parameters stored in memory. The objective is to partially recover the bits of the parameters to efficiently train a substitute model that mimics the victim model with high fidelity.}
  \label{fig:teaser}
\end{figure}

Our method relies on three steps, as illustrated in Fig.\ref{fig:teaser}:
\begin{enumerate}
    \item The adversary builds an attack dataset from pure random inputs with a black-box genetic algorithm. The goal is to feed the victim model with inputs that enhance the efficiency of the safe-error attack.
    \item For each bit of each parameter and for each input from the attack set, the adversary collects two prediction sets: an \textit{error-free} one with the nominal victim model and a \textit{faulted} one with the faulted model. The fault correspond to a bit-set performed with an injection means on the victim model stored in memory. Then, a safe-error attack is performed by comparing the two prediction sets. Non-similar prediction scores enable to recover a \texttt{0} bit value. Additional bits can be recovered with a simple heuristic in the case of two non adjacent bits extracted by SEA.
    \item A substitute model is built and trained with a very limited part (8\%) of the victim training set. Training is constrained with the recovered bits with a \textit{mean clustering training} as proposed in~\cite{deepsteal2022}.   
\end{enumerate}

\subsection{Exploiting safe-error attack}
\label{sea}

We performed a SEA on the victim model parameters $W$ in an iterative way to test all its bits. Our fault model is data-dependent since it is a requirement of SEA. Such a bit-set fault model is achievable for LFI in the Flash memory of a microcontroller platform. Hence, we considered this fault model in our work, however all our results can be extended to the bit-reset case, a model that is also sometime encountered for some Flash memories. 

Over a set of test inputs $X$, the SEA relies on the direct comparison between the predictions $Y$ of the victim model $M_{W}$ and the ones $\widetilde{Y}$ output by the faulted victim model $M_{\widetilde{W}}$. 
We check the similarity between $Y$ and $\widetilde{Y}$, noted $S(Y,\widetilde{Y})$. $S(Y,\widetilde{Y})$ is \texttt{True} if we have a strict equality between the two score matrix, \texttt{False} otherwise. 
In the bit-set context, 3 cases are possible as summarized in Tab.~\ref{Truth_table_dfa}. 
When a fault occurs on a bit having \texttt{0} as value, the fault may lead to a different set of  predictions than the normal behavior, therefore the adversary may conclude that the bit is \texttt{0}. If the bit-set has no impact on the predictions \textit{or} if the bit is already set to \texttt{1}, then the adversary cannot recover the bit value. 
Thus, an adversarial objective is to optimize the error propagation when a fault occurs ($\texttt{0}\rightarrow\texttt{1}$) so that a fault is likely to lead to a difference between $Y$ and $\widetilde{Y}$. We analyze how the selection of the inputs can be leveraged by an adversary to reach this objective in the following Section.

\begin{table}[h!]
\begin{minipage}{0.49\textwidth}
\caption{Truth table of SEA}
\begin{center}
\begin{tabular}{ccc|c}
\toprule
$b_{i}$ & $\widetilde{b_{i}}$ & $S(Y,\widetilde{Y})$ & $b_{i}$ estimation\\
\midrule
0 & 1 & False & 0\\
0 & 1 & True & doubt \\
1 & 1 & True & doubt \\
\midrule
\end{tabular}
\label{Truth_table_dfa}
\end{center}
\end{minipage}
\begin{minipage}{0.49\textwidth}
\caption{Accuracy (\%) according to inputs category}
\begin{center}
\begin{tabular}{cccc}
\toprule
Model & All & Uncertain & Certain \\
\midrule
CNN & 79.4 & 57.07 & 92.23 \\
MLP & 94.94 & 68.07 & 98.2 \\
\bottomrule
\end{tabular}
\label{Models_accuracy_according_inputs}
\end{center}
\end{minipage}
\end{table}

\subsection{Efficiency of task-specific inputs}
\label{sorting_inputs}

As a first analysis, we used inputs from test sets of MNIST and Cifar-10 as attack set for the SEA. Interestingly, we observed a significant heterogeneity of inputs on their efficiency to recover bit-level information: for some inputs, bit-set faults do not alter prediction scores whereas other inputs lead to strong alterations. Experimentally, we distinguished two categories according to their prediction scores. A first class (hereafter called ``Certain``) represents inputs that lead to predictions with a single label having the maximum score of 127 (e.g., $[0, 0, 0, 0, 0, 0, 0, 0, 0, 127]$ for 10 classes). The second class (hereafter called ``Uncertain``) gathers the prediction scores with at least two labels having a non-null score (e.g., $[0, 13, 0, 0, 0, 0, 0, 4, 0, 110]$). Table~\ref{Models_accuracy_according_inputs} provides the accuracy of the MLP and CNN models according to these categories. Unsurprisingly, because models reach a strong confidence on the inputs from the Certain class, the accuracy is naturally high, above 90\% on Cifar-10 and close to 100\% on MNIST. Therefore, most of the mispredictions are concentrated in the Uncertain class.

\begin{figure}[h!]
     \centering
     \begin{subfigure}[l]{0.49\linewidth}
         \centering
         \includegraphics[width=0.90\textwidth]{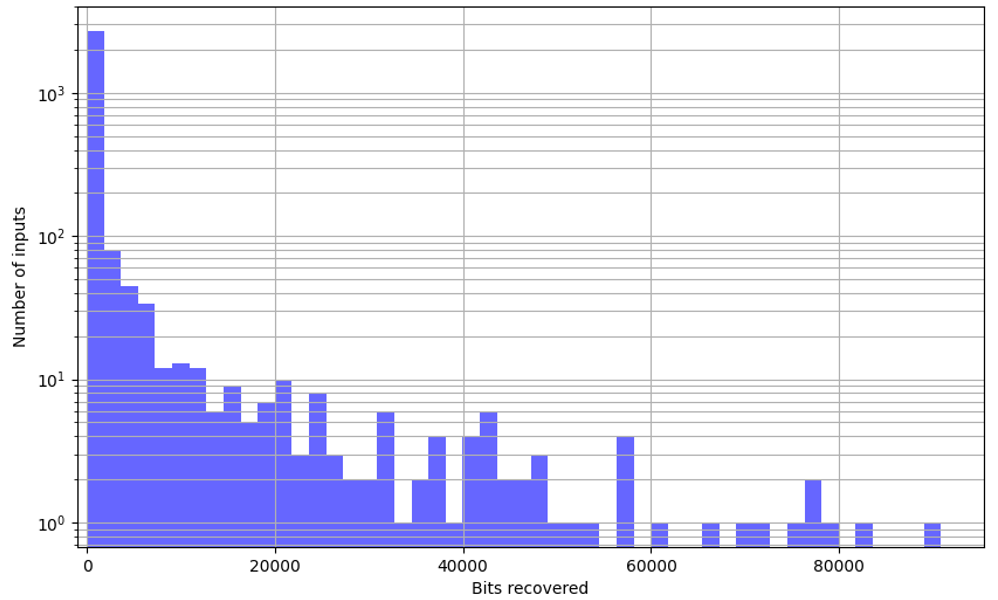}
         \caption{CNN with \textit{Certain} inputs}
         \label{fig:cnn_certains}
     \end{subfigure}
     \hfill
     \begin{subfigure}[r]{0.49\linewidth}
         \centering
         \includegraphics[width=0.9\textwidth]{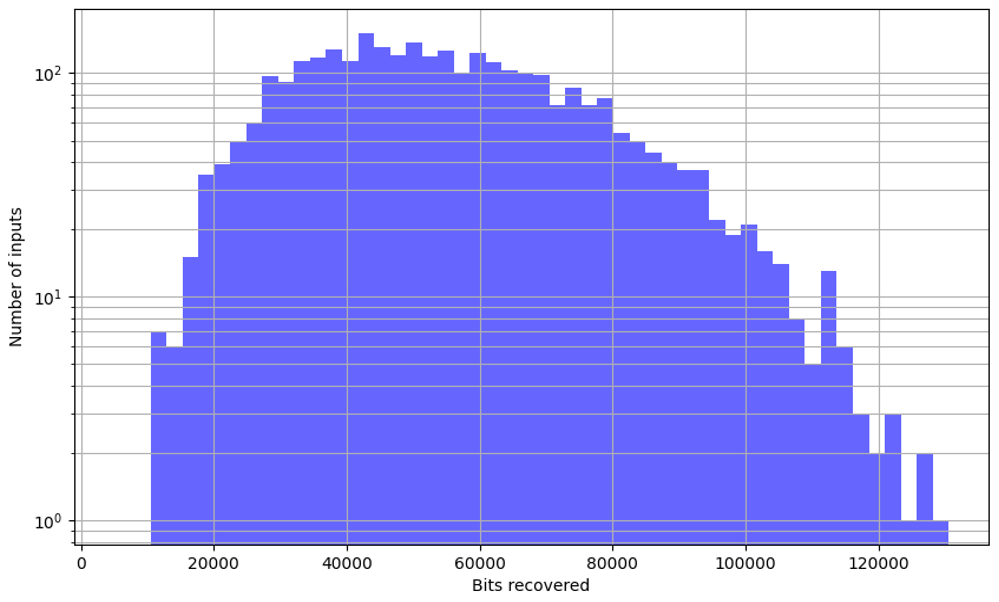}
         \caption{CNN with \textit{Uncertain} inputs}
         \label{fig:cnn_uncertains}
     \end{subfigure}
     \begin{subfigure}[l]{0.49\linewidth}
         \centering
         \includegraphics[width=0.9\textwidth]{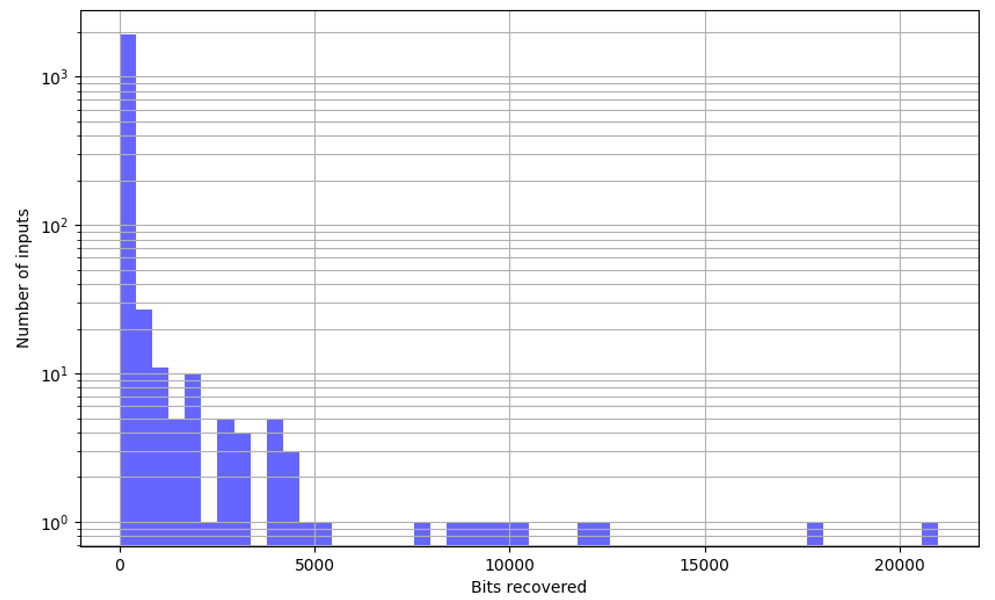}
         \caption{MLP with \textit{Certain} inputs}
         \label{fig:mlp_certains}
     \end{subfigure}
     \hfill
     \begin{subfigure}[r]{0.49\linewidth}
         \centering
         \includegraphics[width=0.9\textwidth]{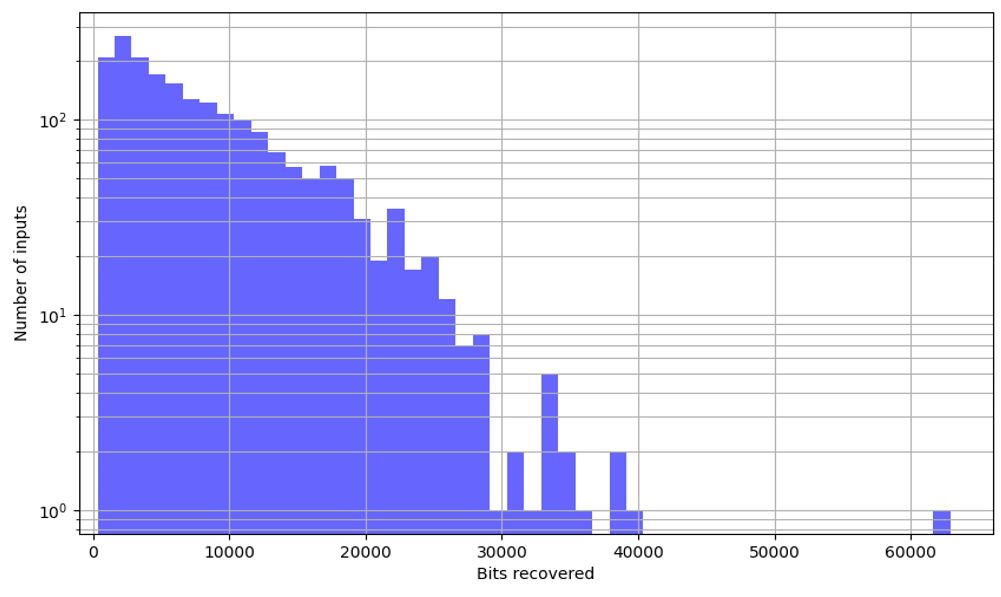}
         \caption{MLP with \textit{Uncertain} inputs}
         \label{fig:mlp_uncertains}
     \end{subfigure}
        \caption{Distribution of recovered bits w.r.t. number of inputs.}
        \label{fig:distribution}
\end{figure}

Tab.~\ref{tab_prediction_global_mnist_cifar10} presents the difference between these two categories of inputs (details per layer are in Appendix, Tab.~\ref{tab_prediction_mnist_cifar10}). The first column is focused on the proportion (\%) of Certain and Uncertain inputs that do not lead to any recovered bit. For example, for CNN, 14.23\% of Certain inputs do not lead to any bit recovery (i.e., predictions between the error-free and the faulted models are identical on these inputs). The last column gives the number of bits recovered over the inputs: for a model, we used $m$ Certain and $m$ Uncertain inputs and computed the average number of bits recovered (and standard deviation) over these $m$ inputs. We used $m=2000$ for MLP and $m=3000$ for CNN. For example, for MLP, inputs from the Uncertain category enable to extract (on average) 64 times more bits than ones from the Certain class (8438 vs. 131).  Fig.~\ref{fig:distribution} (Appendix) shows the distribution of recovered bits w.r.t. the inputs with some outliers that explained high std values.

A first observation is that the Uncertain inputs always leak bit information and the \textit{useless} ones are over-represented in the Certain category for some layers of the two models. This result is more disparate for the CNN model since the parameters of the first convolutional layer are more easily recovered whatever the type of inputs (only 14.33\% of the Certain predictions are useless for the SEA). Moreover, the difference between both categories on the number of recovered bits is significantly higher for the inputs in the Uncertain category, with a factor of 33 for CNN and 64 for MLP.  

\begin{table}[t!]
\begin{minipage}{0.63\textwidth}

\caption{Bit-recovery efficiency for Certain (\textbf{C}) and Uncertain \textbf{U} \\ test inputs.}
\begin{center}
\begin{tabular}{ccccc}
\toprule
\multirow{2}{*}{} & \multicolumn{2}{c}{No recovery (\%)} & \multicolumn{2}{c}{Average \# of} \\

\multirow{2}{*}{} & \multicolumn{2}{c}{} & \multicolumn{2}{c}{bits recovered (std)} \\

& C & U & C & U \\

\midrule
MLP  & 47.65 & 0.0 & \textbf{131} (969) & \textbf{8438} (6856) \\
\midrule
CNN  & 14.23 & 0.0 & \textbf{1656} (7066) & \textbf{55388} (21229) \\
\bottomrule

\end{tabular}
\label{tab_prediction_global_mnist_cifar10}
\end{center}

\end{minipage}
\begin{minipage}{0.35\textwidth}

\caption{Prediction types over 5000 random inputs.}
\begin{center}
\begin{tabular}{lrr}
\toprule
 & MLP & CNN \\
\midrule
C (\%) & 9.82 & 99.4\\
U (\%) & 90.18 & 0.06\\

\bottomrule
\end{tabular}
\label{Random_inputs_efficiency_against_MLP_and_CNN}
\end{center}

\end{minipage}
\end{table}

\begin{figure}[b!]
     \centering
     \begin{subfigure}[l]{0.49\linewidth}
         \centering
         \includegraphics[width=0.95\textwidth]{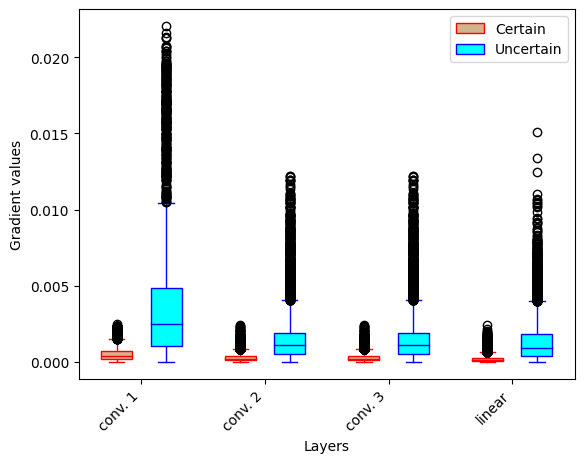}
         \caption{CNN}
         \label{fig:gradients_cnn}
     \end{subfigure}
     \hfill
     \begin{subfigure}[r]{0.49\linewidth}
         \centering
         \includegraphics[width=0.95\textwidth]{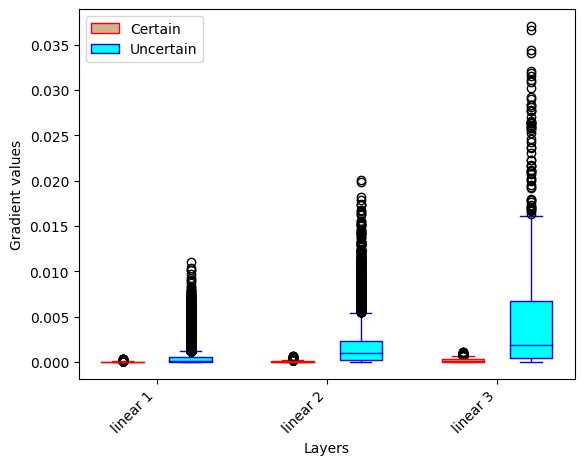}
         \caption{MLP}
         \label{fig:gradients_mlp}
     \end{subfigure}
        \caption{$\nabla_{w}\mathcal{L}$ distribution per layer for Certain and Uncertain inputs. The boxplot represents the median value inside the first and third quartiles. Blue lines extend the box by 1.5x and black circles are outliers.}
        \label{fig:gradients_cnn_mlp}
\end{figure}

We propose a closer look of this phenomenon by analyzing the distribution (per layer) of the absolute  gradient values of the loss w.r.t. the parameters ($\nabla_W\mathcal{L}_{CE}$) represented in Fig.~\ref{fig:gradients_cnn_mlp}. We observe a clear difference of the loss sensitivity between the two categories with high magnitudes for the Uncertain class: for inputs leading to uncertain predictions, a modification of the parameters will strongly affect the loss value and, therefore, the predictions.

This first experiment demonstrates the strong impact of input nature on the exploitation of a SEA to recover bit values. Setting in a threat model where the adversary has a very limited access to training dataset, these results pave the way to a strategy that aims at crafting inputs with a single objective: provide \textit{uncertain} predictions. That point is important since the challenge is less complex than the generation of task-specific inputs similar to ones belonging into the original training distribution.

    \subsection{Crafting inputs for uncertain predictions}
    \label{crafing_inputs}
    \noindent\textbf{Random inputs. }
Our first interrogation concerns the efficiency of random inputs. With a very limited access to the training dataset, the use of unlimited random inputs could be a real advantage for the bits recovery step. Tab.~\ref{Random_inputs_efficiency_against_MLP_and_CNN} shows the dispatching between the two prediction categories of 5000 random inputs following a uniform distribution. For MLP, random inputs allow to generate an attack set with 90,18\% of inputs leading to Uncertain predictions. Surprisingly, for CNN, only Certain predictions are obtained. We noticed that the random inputs are (nearly) always associated to the same label (label \texttt{7}), therefore all the predictions are equal. We observed the same behavior for another CNN architectures trained on Cifar-10 (VGG-8 with or without biases).
Since architecture of the victim model is an important factor on the efficiency of random inputs, we suggest a black-box approach to craft inputs leading to Uncertain predictions when inferred with the model thanks to a genetic algorithm.
\newline
\newline
\textbf{A black-box crafting method. }Our goal is to use a genetic algorithm (GA) to directly craft an attack set with as only objective to produce Uncertain predictions. Our GA acts as follow. 

\textit{(1) Population initialization.} GA starts with a set of inputs, called \textit{population}, that is simply sampled using a uniform distribution (as for the random inputs above) with pixel values in $[V_{min}, V_{max}]$ such as $[V_{min}, V_{max}] \subset [-127;127]$. We constrained the range value since we experimentally observed that it helps the algorithm convergence. Several values of $(V_{min},V_{max})$ can be fixed to have several initial population and increase the diversity. In our case 150 elements are used per population.

\textit{(2) Objective definition.} GA aims at performing iterative transformations on the population to reach our adversarial objective: the output scores must match with an uncertain prediction scores fixed by the adversary, called \textit{target scores} $Y^t$. A target score, $y^t$ is in $\mathbb{R}^{K}$ with $y^t(i) \in [0;127]$. $C$ scores among $K$ are randomly picked and set to $0$, with $C \in [0;N-2]$. The $K-C$ non-null remaining scores are randomly set and scaled so that $\sum_{i}y^t(i) = 127$.

\textit{(3) Population evolution.} The core process is iterative and aims at building a new population thanks to a set of classical transformations. At iteration $t$, the new population results from \textit{selection}, \textit{crossover} and \textit{mutation} operations between elements of the previous population at $t-1$. A new population is generated until an optimal solution is reached. To determine which operation is done on each element of the population a cost function is used in order to sort elements of the population. In our case, the cost function is $\mathcal{L}_{CE}$. The \textit{selection}  consists in keeping some elements of the previous generation at $t-1$ to create the new one. The \textit{selection} keeps $b$\% of the best elements and $r$\% of the sub-optimal elements, randomly chosen (typical values for CNN are $b=60$ and $r=20$). The \textit{mutation} operation  randomly applies some noise in sub-optimal elements of the previous generation to create new elements. The \textit{crossover} merges two elements of the previous generation to create two new elements by interchanging half of each element randomly. 

GA based  method is repeated until the number of elements wanted in the attack dataset is reached. In Fig. \ref{fig:SEA_LLSB_results} (left), we observe on CNN that our GA-generated inputs are significantly more efficient than random inputs (dotted lines). We extract 80\% and 90\% of the MSB of the CNN parameters with only 150 and 1500 inputs, respectively. 

    \subsection{Least Significant Bit Leakage Principle}
    \label{bits_estimation}
    \begin{figure}[t!]
\centerline{\includegraphics[width=0.70\textwidth]{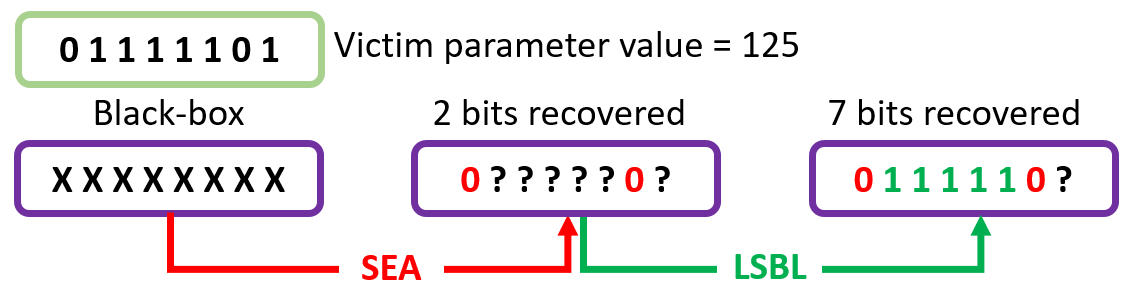}}
\caption{Illustration of the LSBL principle.}
\label{fig:llsb_explication}
\end{figure}

Because of the bit-set fault model and its intrinsic ambiguity (cf. truth table~\ref{Truth_table_dfa}), SEA only enables to \textit{partially} recover the bit values of the victim model parameters. However, it is possible to increase the number of bits recovered by applying a principle that we called the \textit{Least Significant Bit Leakage} principle (hereafter, LSBL) that enables to guess bit values according to the position of the bits already recovered.

The LSBL principle is as follows: \textit{if a bit $b_{k}$ with $k \in ]0;7]$ of a parameter $w$ has been recovered by SEA (i.e., $b_{k}=\texttt{0}$ without ambiguity), then all undefined bits $b_{i}$ with $i \in [0;k-1]$ can be estimated to \texttt{1}.} 

An explanation is that the more a bit-set is performed on a least significant bit than $k$, the smaller the variation of the parameter is. Thus, if a small alteration impacts prediction (at $k$) then, a bigger alteration should also impact it. This principle enables to estimate the bits $b_{i}$ because if these bits do not impact the prediction, it is due to a bit value equal to \texttt{1}. Fig.~\ref{fig:llsb_explication} illustrates LSBL with one parameter $w=125$. With SEA, we recover $b_{0}$ and $b_{6}$. Applying LSBL, we can guess all undefined bits from $b_1$ to $b_5$. Thus, LSBL enables to grow the extracted information from 2 to 6 bits. Thanks to the LSBL principle the rate of recovered bits increase from 47.05\% to 80.1\% for the CNN model with 5000 crafted inputs.

\begin{figure*}[h!]
\centering
\includegraphics[width=\textwidth]{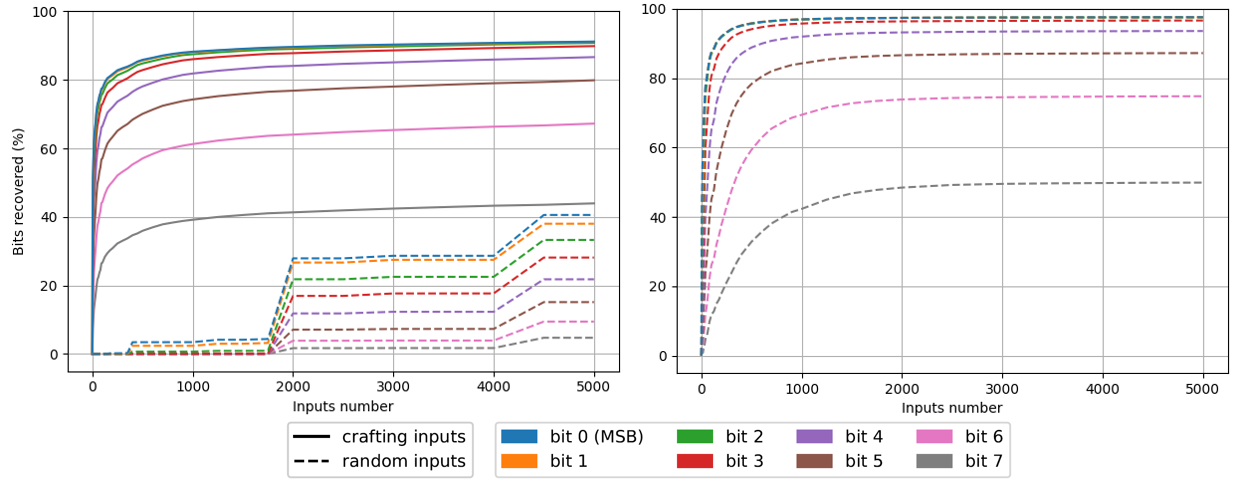}
\caption{Bits recovered with SEA and LSBL for (left) CNN and (right) MLP (random inputs only).}
\label{fig:SEA_LLSB_results}
\end{figure*}

In Fig.~\ref{fig:SEA_LLSB_results} we shows the percentage of bits recovered by combining SEA and LSBL according to the inputs. We propose both random and GA-crafted inputs for the CNN model. To ease the visualization, we only present random inputs for MLP, since GA provide similar results (see ~\ref{crafing_inputs}). 
Our results demonstrate a high rate of recovered bits for both models thanks to SEA associated to the LSBL principle. In the best case, we can estimate about 90\% of the most significant bits. Our method is also efficient for the 6 MSBs with a recovery rate superior to 80\%. Moreover, we notice a fast plateau effect with the majority (80\%) of the recovered bits extracted with approximately 250 inputs (CNN) then, after 1500 inputs (CNN), using more inputs only allow to recover few bits. The same effect is observed for MLP. We evaluated the LSBL principle on CNN and MLP models by computing the recovery error for the bits exclusively estimated by LSBL. The recovery error rapidly goes under 1\% for only 150 and 300 inputs for CNN and MLP respectively. Therefore, this heuristic, despite being perfect, enables to recover bit values with a very low error rate.  

    \subsection{Train substitute model}
    \label{target_training}
    As in~\cite{deepsteal2022}
, we trained the substitute model by using the recovered bits as a constraint over only 8\% of the training dataset. Without training (i.e. if we simply set the recovered bits to their estimated values and randomly initialized the other bits) we reached a very low performance of the substitute model with an average accuracy of 26.02\% for CNN and 75.78\% for MLP on the test set with at least 90\% of MSB.
Similarly as \cite{deepsteal2022}, we used a new loss (hereafter noted $\mathcal{L}^{sub}$, for the \textit{substitute} loss) that relies on the cross-entropy loss $\mathcal{L}_{CE}$ (so that the substitute model is trained to perform the same task-oriented objective as the victim) with a penalty term that constrains the partially recovered weights. As in~\cite{deepsteal2022}, $\mathcal{L}^{sub}$ is defined as (Eq.~\ref{eq:eq_training_loss}): 

\begin{equation}
\mathcal{L}^{sub}=\mathcal{L}_{CE}\big( M'_{\Theta}(x),y\big) + \lambda \sum_{l=1}^{L}||\Theta^l-\Theta^l_{mean}||
\label{eq:eq_training_loss}
\end{equation}

with $L$ the number of layers, $\Theta^l$ is parameters matrix of layer $l$ of the substitute model $M'$, $\Theta^l_{mean}=(\Theta^l_{min} + \Theta^l_{max})/2$ are the mean values according to the \textit{min} and \textit{max} values, updated after each training iteration and $\lambda$ a hyper-parameter balancing the penalty strength. Initial values are computed with the MSB recovered that define \textit{projected ranges} for the possible values of the parameters partially extracted. For example, if the two first MSB of  $\theta$ are \texttt{0}, then the projected range for $\theta$ is $[0;63]$. For each training batch, the mean clustering updates $\Theta_{mean}$ with the current values and parameters are clipped according to the projected range after each training epoch. The main objective is to avoid any divergence of parameters from the information extracted by the fault injection step. Note that this training procedure is only applied to parameters partially recovered by SEA. For parameters without any recovered bit, no penalty can be applied, therefore $\lambda=0$. For parameters fully recovered (i.e., the 8 bits), they are freezed at training time. As demonstrated in~\cite{deepsteal2022}, minimizing $\mathcal{L}^{sub}$ with the mean clustering training allows to train the substitute model with few training data.

    \subsection{Evaluation}
    \label{evaluation}
    We keep the same evaluation protocol as in~\cite{deepsteal2022}. A first criteria is the \textit{accuracy} reached by the substitute model after training when tested with test set of MNIST (for MLP) and Cifar-10 (for CNN). The closer to the victim performance the better. The second criteria is the \textit{fidelity} between substitute and victim models defined as the rate of identical predictions over test set (the higher the better). The last criteria is \textit{Accuracy Under Attack} (AUA) which is the accuracy of victim model when fed with adversarial examples crafted on substitute model. For AUA, if substitute model achieves to mimic victim model behaviours, then both models will respond similarly when facing adversarial examples: the transferability of adversarial perturbations will be maximum and we expect a very low adversarial accuracy for victim \textit{and} substitute model. AUA is twice interesting since it measures a similarity-level between two models as well as the capacity of the adversary to craft efficient adversarial examples against victim model thanks to his substitute model. Consistently with the state-of-the-art, we used the $l_\infty$-PGD attack~\cite{madrytowards} with 40 steps and an adversarial budget of $\epsilon=8/255$ for Cifar-10 and $\epsilon=0.3$ for MNIST.

\begin{table*}[t!]
\caption{Performance of the substitute model after training according to the level of extraction. (AUA: Accuracy Under Attack)}
\begin{center}
\begin{tabular}{ccccccc}
\toprule
At least \% MSB recovered & \multicolumn{3}{c}{CNN} & \multicolumn{3}{c}{MLP}  \\
(+ others if recovered)&  Accuracy & Fidelity & AUA & Accuracy & Fidelity & AUA \\

\midrule
90 &  75.27 & 85.58 & 1.83 & 92.93 & 96.44 & 0.0 \\
80 &  69.36 & 77.00 & 5.55 & 92.09 & 95.48 & 0.01 \\
70 &  54.59 & 61.10 & 12.99 & 90.52 & 93.66 & 0.1 \\
60 &  40.55 & 44.66 & 34.84 & 64.50 & 66.56 & 12.50 \\
\midrule
\textit{Victim} & 79.4 & 100 & 0.42 & 94.94 & 100 & 0.0 \\
\bottomrule
\end{tabular}
\label{tab_strong_unsure_cnn_cifar10}
\end{center}
\end{table*}

Our results are summarized in Tab.~\ref{tab_strong_unsure_cnn_cifar10} for MLP and CNN according to different recovery rate of MSB. For CNN, accuracy grows from 40.55\% to 75.27\% when MSB recovered ratio increases from 60\% to 90\% (64.50\% to 92.93\% as accuracy for MLP). Our best results for CNN and MLP represent an accuracy drop of only 4\% and 2\% respectively compared to victim models. Importantly, we also reach high fidelity rate with 85.58\% (CNN) and 96.44\% (MLP) for the best case (90\% of MSB). Focusing on the CNN results, the performance of our approach is consistent with the ones observed in~\cite{deepsteal2022} with different architectures\footnote{On Cifar-10, with 90\% of recovered MSB, \cite{deepsteal2022} built a ResNet-18 substitute model with an accuracy of 89.59\% (victim:93.16\%), a fidelity of 91.6\% and AUA of 1.61 (victim: 0\%), and a VGG-11 substitute model with an accuracy of 81.56\% (victim:89.96\%), a fidelity of 83.33\% and AUA of 18.55\% (victim: 4.63\%).}. AUA results show that adversarial examples can be efficiently crafted from substitute models, then applied on the victim models. With the best amount of MSB recovered, victim models have an adversarial accuracy close to the one obtained by crafting adversarial examples in a white-box context (1.83\% for CNN and 0\% for MLP). Generally, whatever numbers of MSB used to train substitute models, adversarial examples crafted on substitute models are sufficiently efficient on victim models to have an adversarial accuracy below 35\%.

\section{Discussions}
\label{discussion}
\label{limitations_and_further_analysis}

\noindent\textbf{Impact of model architecture.} We evaluated our method on two classical models relevant for the type of platforms we consider. However, our results raise open questions about the impact of the model architecture on the SEA efficiency. Further analysis have to be investigated on other models and layers, e.g. influence of the model depth, residual networks (ResNet) or batch normalization. 

\noindent\textbf{Input generation strategy.}
A limitation of our GA-based crafting method is the use of too many parameters at different steps of the process (objective definition, inputs initialization and transformations applied to the population). Therefore, further analysis should evaluate alternative techniques such as the Black-Box Ripper approach proposed by Barbalau \textit{et al.} \cite{barbalau2020black} that exploit generative evolutionary framework to build a proxy dataset. Other methods may rely on black-box decision-based adversarial examples~\cite{brendel2018decision} to craft Uncertain samples. 

\noindent\textbf{Practical LFI experiments.}  
Colombier \textit{et al.}~\cite{colombier2019laser} first demonstrated the bit-set fault model in Flash memory of a Cortex-M 32-bit microcontrollers with LFI. It was directly applied in \cite{dumont2023evaluation} to evaluate the robustness of an embedded neural network against a weight-based adversarial attack with a bit-set variant of the Bit-Flip Attack from~\cite{rakin2019bit}. To evaluate the practicability of our method, we strictly followed the set up from \cite{dumont2023evaluation} and conducted first experiments with a MLP model on an ARM Cortex-M3 platform. The model is the same as in \cite{dumont2023evaluation} with two linear layers (50 - 10 neurons) trained on MNIST with dimensionality reduction on $\mathbb{R}^{50}$ performed by principal component analysis. The fault injections setup is composed of a two-spot laser beam in near infrared (Cf. Set up details in Appendix).
As a first practical experiment and for characterization purpose, we set in a white-box context with trigger signal to monitor the synchronization of the laser beam with the loading of the targeted parameter. Moreover the inference program has been compiled without optimization (\texttt{O0}). We recovered 90\% of MSB (with other bits if recovered) by using only 15 crafted inputs. As mentioned in~\cite{colombier2019laser} and \cite{dumont2023evaluation}, we noted that the bits can be targeted very precisely with a perfect repeatability of the fault injection.  

The main limitation in terms of practicability is that the attack needs one inference with one fault injection per bit and per input (attack set). Then, the overall SEA time is $T_{SEA} = N_{inputs}\times N_{bits} \times \delta_{inf}$ with $\delta_{inf}$ the inference time. $\delta_{inf}$ could be a real bottleneck for constrained Cortex-M platforms. Without any optimisation (e.g., compilation level) we had $\delta_{inf}=150$ ms, then the complete non-optimized attack lasted 3 hours. However, targeting every bit of the model is not necessary and the attack duration can be reduced, at least, by half by considering the 4 MSB or even 20 minutes with only the first MSB as in~\cite{deepsteal2022}. Importantly, we highlight the fact that complexity comparison with~\cite{deepsteal2022} is hardly possible since DRAM-CPU platforms in~\cite{deepsteal2022} can handle far more complex models than 32-bit microcontroller devices as in our work\footnote{In~\cite{deepsteal2022}, authors used ResNet-18,-34 and VGG-11 on Cifar-10 with an average extraction time of 12 days (extraction of the first MSB)}.

This work paves the way to further researches that should be focused on combining side-channel analysis to trigger the faults in black-box setting as well as using different 32-bit microcontroller platforms. Moreover, other injection mean should be studied, such as electromagnetic fault injection (EMFI) since bit-reset fault models have been demonstrated even in CPU platforms \cite{trouchkine2021fault}.

\noindent\textbf{How to protect?} Protections against fault injections and safe-error attacks classically encompass randomization,  redundancy and data integrity check (error-correction). However, most of the traditional defenses can be too expensive to protect a whole model (i.e. all bits). A logical way to overcome our attack is to add randomization within the model so that it slightly perturbs the prediction scores without altering the overall performance of the model. This can be achieved by randomly scaling the output feature map of the intermediate layers of the model. We test our protection on the last convolutional layer of our CNN by used 8 scaling factors $\alpha_i$ that are randomly picked in $[0.9, 1]$ at each inference (each $\alpha$ scales 8 channels). With this strategy, we keep an overall performance with a accuracy that slightly drops from 79.40\% (nominal accuracy) to 79.30\% and our SEA approach is no more able to extract bit values because the predictions are unlikely to be equal from one inference to another. 
 
The main limitation on such randomization approach is that it does not make the attack impractical since the adversary may rely his strategy on an expectation of the predictions to drown the effect of the randomization. In Tab.~\ref{protection_evaluation}, we represent the average and standard deviation of the difference between two groups of prediction scores where each input is repeated $N$ times (i.e., the output score for this input is averaged over $N$ inferences, then the difference is averaged over the 10 labels). For convenience, this difference is simply noted as $\Delta Y$. As expected, we observe that the impact of the randomization is significantly drown as $N$ grows. However, this theoretical limitation needs to be moderated with a practical point of view since we deal with an attack that relies on fault injections. Indeed, this first result shows that the adversary would need to process at least 1000 inferences (hence 1000 faults) for each input to thwart the protection. Such a drastic practical overload could be prohibitive for many fault injection means.

\begin{table}[t!]
\caption{Impact of expectation over layer randomization. Average (std) of the scores between two set of predictions for CNN (5000 crafted inputs).}
\begin{center}
\begin{tabular}{ccccc}
\toprule
&\multicolumn{4}{c}{Expectation over $N$ inferences per input} \\
N= & 2 & 10 & 100 & 1000 \\
\midrule
$\Delta Y$ (std) & 7.7 (12.3) & 2.4 (3.7) & 0.8 (1.3) & 0.3 (0.6) \\
\bottomrule
\end{tabular}
\label{protection_evaluation}
\end{center}
\end{table}

\section{Conclusion}
Our model extraction attack is specifically adapted to models deployed in constrained platforms that are vulnerable to memory alterations. We extract information from a victim model by using a safe-error attack principle with custom inputs that optimize the leakage of parameters with bit-set fault injections. We present promising results on two architectures (MLP and CNN) with a successful extraction of 80\% of the 6 most significant bits of victim parameters. These recovered bit values are used to constrain the substitute model training even with very limited training data that finally reaches similar accuracy than the victim model with a high-level of fidelity. This work aims at highlighting the criticity of model extraction regarding the large-scale deployment of machine learning models in hardware platforms. Our work paves the way to further practical experiments with different fault injection means and target devices as well as suggestion of potential protections.

\section*{Acknowledgment}
This work is supported by (CEA-Leti) the European project InSecTT\footnote{\url{www.insectt.eu}, ECSEL JU 876038} and by the French ANR in the \textit{Investissements d’avenir} program (ANR-10-AIRT-05, irtnanoelec);  and (MSE) by the ANR PICTURE program\footnote{\url{https://picture-anr.cea.fr}}. This work benefited from the French Jean Zay supercomputer with the AI dynamic access program.

\bibliographystyle{splncs04}
\bibliography{biblio/biblio}

\section*{Appendix}

\subsection*{Details of the Bit-recovery efficiency}
\begin{table}[h!]
\caption{Bit-recovery efficiency for \textit{Certain} and \textit{Uncertain} test inputs. (top) MLP and (bottom) CNN.}
\begin{center}
\resizebox{0.65\textwidth}{!}{%
\begin{tabular}{ccccc}
\toprule
\multirow{2}{*}{Layers} & \multicolumn{2}{c}{No recovery (\%)} & \multicolumn{2}{c}{Average \# of bits recovered (std)} \\
& Certain & Uncertain & Certain & Uncertain \\
\midrule
Linear 1 & 97.15 & 0.0 & 83 (915) & 5976 (5564) \\
Linear 2 & 92.6 & 0.0 & 40 (301) & 2229 (1478) \\
Linear 3 & 48.95 & 0.0 & 8 (21) & 231 (87) \\
\midrule
MLP  & 47.65 & 0.0 & 131 (969) & 8438 (6856) \\
\midrule
\midrule
Conv. 1 & 14.33 & 0.0 & 399 (933) & 7144 (1309) \\
Conv. 2 & 68.6 & 0.0 & 895 (4255) & 32380 (11635) \\
Conv. 3 & 79.5 & 0.0 & 339 (2057) & 15062 (9463) \\
Linear 1 & 67.5 & 0.0 & 24 (87) & 803 (406) \\
\midrule
CNN  & 14.23 & 0.0 & 1656 (7066) & 55388 (21229) \\
\bottomrule
\end{tabular}
}
\label{tab_prediction_mnist_cifar10}
\end{center}
\end{table}

\subsection*{LFI experiments}

\noindent Target board: 
\begin{itemize}
    \item ARM Cortex-M3, 8\,MHz ($90$\,nm CMOS), 128\,kB Flash memory.
    \item MCU packaging was opened with engraving.
    \item Communication with ChipWhisperer CW308 platform.
\end{itemize}

\noindent LFI platform: 
\begin{itemize}
    \item 2 spots near infrared, $\lambda=$ $1,064$\,nm, spot diameter: [$1.5$, $15$]\,µm, maximum power: $1,700$\,mW.
    \item Experience: power=$170$\,mW, pulse width = $200$\,ns, lens magnification $\times5$, spot diameter = $15$\,µm
\end{itemize}

\subsection*{Distribution of recovered bits}

\end{document}